\documentclass{article}

\usepackage{rotating}
\usepackage{natbib}

\begin{document}

\title{Adjusting for selective non-participation with re-contact data in the FINRISK 2012 survey}

\author{Juho Kopra$^{1}$, Tommi H{\"a}rk{\"a}nen$^{2}$, Hanna Tolonen$^{2}$, Pekka Jousilahti$^{2}$,\\ Kari Kuulasmaa$^{2}$, Jaakko Reinikainen$^{2}$ and Juha Karvanen$^{1}$\\[1cm]
	$^{1}$
	Department of Mathematics and Statistics,\\ University of Jyvaskyla, Finland\\
	$^{2}$
	Department of Public Health Solutions,\\ National Institute for Health and Welfare, Helsinki, Finland\\[1cm]
	Corresponding author:\\
	~~Juho Kopra\\
	~~Department of Mathematics and Statistics\\
	~~P.O. Box 35 (MaD)\\
	~~FI-40014 University of Jyv\"askyl\"a\\
	~~Finland\\
	~~Email: juho.j.kopra@jyu.fi
}
\date{March 23, 2017 (submitted version)}

\maketitle

\begin{abstract}
	\textbf{Aims:} A common objective of epidemiological surveys is to provide population-level estimates of health indicators. 
	Survey results tend to be biased under selective non-participation. 
	One approach to bias reduction is to collect information about non-participants by contacting them again and asking them to fill in a questionnaire.
	This information is called re-contact data, and it allows to adjust the estimates for non-participation.
	
	\textbf{Methods:} We analyse data from the FINRISK 2012 survey, where re-contact data were collected. 
	We assume that the respondents of the re-contact survey are similar to the remaining non-participants with respect to the health given their available background information. Validity of this assumption is evaluated based on the hospitalization data obtained through record linkage of survey data to the administrative registers. 
	Using this assumption and multiple imputation, we estimate the prevalences of daily smoking and heavy alcohol consumption and compare them to estimates obtained with a commonly used assumption that the participants represent the entire target group.
	
	\textbf{Results:} This approach produces higher prevalence estimates than what is estimated from participants only. Among men, smoking prevalence estimate was $28.5\%$ ($23.2\%$ for participants), heavy alcohol consumption prevalence was $9.4\%$ ($6.8\%$ for participants). Among women, smoking prevalence was $19.0\%$ ($16.5\%$ for participants) and heavy alcohol consumption $4.8\%$ ($3.0\%$ for participants).
	
	\textbf{Conclusion:} Utilization of re-contact data is a useful method to adjust for non-participation bias on population estimates in epidemiological surveys.
\end{abstract}
%

\section{Introduction}
Health examination surveys (HES) are among the key data sources for data-driven planning of national health policies. If the participants of the survey are a representative sample of the population of interest, then simple statistical estimates, such as sample averages, provide reliable support to decision-making. 
A major threat for representativeness of survey data is selective non-participation. 
Under selective non-participation, the survey participant data do not represent the population of interest, which leads to bias in population-level health indicators. By health indicator we mean a health related population statistic, such as the prevalence of smokers. 
For example, if healthy people are more willing to participate in a survey than people with poor health, the health indicators give an overly positive impression of the health of the population. 
This makes the data misleading. 

The sampling frame often provides background information, such as sex, age and region, on the sample members. This information reveals if some demographic groups are under- or overrepresented among the participants compared to non-participants. However, this information is insufficient to assess whether or not the non-participation is selective with respect to variables of interest.

Record linkage of HES studies with register-based data has shown that non-participants have higher alcohol consumption \citep{torvik2012alcohol,gray2013use}, and higher smoking and alcohol-related mortality \citep{christensen2015wrong}, and higher total mortality rate \citep{jousilahti2005total,thygesen2008effects,larsen2012mortality} than participants. 
In addition, non-participants tend to be younger and less educated than participants. They receive more often social welfare payments and have higher unemployment rate \citep{drivsholm2006representativeness}. 

External information 
are required to assess possible selectivity. 
Two sources of external information considered in this paper are follow-up data and re-contact data.

Follow-up data are (time-to-event) data collected after the survey about disease diagnoses with date details and/or date of death and causes of death. Re-contact data are data from a survey conducted among people who did not participate at the actual survey. 

Recently, adjustment methods using follow-up \citep{kopra2015correcting} and re-contact data \citep{karvanen2016selection} as an additional source of information have been proposed to reduce the selection bias. Kopra et al. \citep{kopra2015correcting} utilized a Bayesian survival model to impute the values of daily smoking using register-based follow-up data on COPD and lung cancer. Karvanen et al. \citep{karvanen2016selection} used data on re-contact respondents information and evaluated the modelling assumption using the five years of register-based follow-up data. A problem with these methods is that they can be applied only after the several years of follow-up after the survey has finished. In this paper, we use the same main idea as in \citep{karvanen2016selection}.  However, an important difference is that we evaluate the modelling assumption using data on past hospitalizations instead from follow-up data.

The aim of this paper is two-fold: 1) To provide estimates for prevalences of self-reported heavy alcohol consumption and daily smoking corrected for non-participation using data obtained through re-contact of non-participants. 2) To use register-based data on hospitalization history for the evaluation of the assumption about similarity of the health of re-contacted non-participants and the rest of the non-participants.

\section{Methods}
\subsection{Data}
We use data from the National FINRISK 2012 survey, a health examination survey (HES) among adults from five regions of Finland \citep{borodulin2013kansallinen}. The survey was conducted in early 2012 with a total sample size of $10,000$ invitees aged 25--74 years. The invitees were sampled from the national Population Information System using simple random sampling stratified by region, sex and 10-year age group. The survey has been approved by the respective Ethics Committee at the time when the survey was conducted. A written informed consent was obtained from survey participants. 

The survey participants filled in a questionnaire at home and participated in a health examination at a local examination clinic, which included physical measurements and collection of biological samples. 
SMS reminders before appointment were used to encourage the invitees to participate.

A total of $5827$ invitees participated in the survey, yielding a $58.3$\% participation rate (i.e. having both the questionnaire and the health examinations completed). Those who did not participate in the health examination were re-contacted by mail ($4173$ in total). The re-contact letter, which should not be confused with the SMS reminders, included identical questionnaire as the initial contact together with pre-paid return envelope and a request to fill-in and return the questionnaire. Thus, the questionnaire and the questions were the same as the original invitation. The re-contact round resulted in $597$ returned questionnaires ($14.3$\% of all non-participants), leaving $3576$ non-participants without any self-reported information. The time lag between the initial survey and re-contact round was 2--5 months. During that time some persons may have changed their smoking or alcohol use habits, but we expect this not to alter the results notably.

The data on background variables, sex, age and region, were available from the sampling frame for both participants and non-participants. 
Data on hospital visits and diagnoses (ICD-codes) since 1969 were obtained for both participants and non-participants through record linkage to the National Hospital Discharge Register \citep{hilmoweb} using the unique personal identification code provided for every resident in Finland. We call these data hospitalization history data.

The survey sample was classified into three groups of people:
\begin{enumerate}
	\item	\underline{Participants} who returned the questionnaire and participated in a health examination.
	\item	\underline{Re-contact respondents} who did not participate in the survey after initial invitation, but did return the re-contact round questionnaire.
	\item	\underline{Non-participants} who neither participated in health examination nor returned the re-contact round questionnaire.
\end{enumerate}

The variables of primary interest are self-reported daily smoking and heavy alcohol consumption. 
Females who consumed more than 16 portions of alcohol per week and men who consumed more than 24 portions per week were defined as heavy alcohol users. One portion corresponds to 12 grams of pure alcohol.

\subsection{Modelling approach}
We fit two kinds of models: three alternative models to impute missing values in data and one model to evaluate the modelling assumptions. 
We apply R statistical software version 3.3.1 \cite{R} and the R-package \texttt{mice} \cite{Rmice} for multiple imputations and R-package \texttt{pscl} for the evaluation of the modelling assumptions.

The alternative modelling assumptions which we consider here, are
\begin{enumerate}
	\item[(1)]	The participants represent the whole population of interest.	
	\item[(2)]	The participants represent the whole population of interest when adjusted for background variables.
	\item[(3)]	The re-contact respondents represent all non-participants when adjusted for background variables.
\end{enumerate}
Assumption (1) is a missing completely at random (MCAR) assumption \citep{littleandrubin} leading to the complete case analysis where data on participants are used to estimate the health indicators of non-participants. 
If assumption (1) holds, the non-participation is neither selective with respect to variables of interest nor background variables. It means that e.g. the average prevalence of smoking measured from the participants describes the smoking prevalence for the whole population even without adjusting for background variables. 
This assumption is made implicitly when estimates based on participants only are reported. 

Assumption (2) is a missing at random (MAR) assumption that makes it possible to use data on participants to estimate the health of non-participants and therefore the health of the whole population provided that the background variables are collected. 
If assumption (2) holds, then non-participation is not selective with respect to variables of interest, but it may be selective with respect to the background variables. To estimate e.g. the prevalence of smoking for the non-participants, adjustment for the background variables is required. 

The assumption (3) allows the use of data on re-contact respondents 
to estimate the health of non-participants provided that background variables are collected for all invitees of a survey. This assumption can be interpreted as a version of the continuum of resistance model \citep{lin1995,boniface2017} where we adjust for background variables. Under assumption (3), the data are missing not at random (MNAR) with respect to HES participation, and MAR with respect to re-contact response.

Under assumption (3), participation may be selective with respect to variables of interest and background variables. However, the response to re-contact questionnaire is not selective with respect to variables of interest, but it may be selective with respect to background information. If this assumption does not hold the health indicators of remaining non-participants cannot be estimated without bias unless some additional data are available.

\subsection{Imputation models}
We consider three different approaches imputing the health indicators. 
The approaches utilize either assumption (2) or (3). Assumption (1) is not used in imputation but is utilized if estimates based on data on participants only are considered to describe the health of the whole population. 
Our main approach is called MI-MNAR, using multiple imputation (MI) with an assumption (3). In MI-MNAR the missing values for re-contact respondents and non-participants are imputed assuming that the parameters of the model are different for re-contact respondents and participants. Two alternative methods use assumption (2), and are called MI-MAR and MI-MAR-NR. In MI-MAR imputation the model parameters are the same in all groups. MI-MAR-NR method uses no re-contact (NR) data at all but is otherwise similar to MI-MAR.

For each imputed variable the multiple imputations are carried out using a regression model (fully conditional specification) \citep{van2012flexible}. The other variables are used as covariates in the imputation model. The imputed variables are daily smoking and heavy alcohol consumption which are predicted by the following covariates: sex, age, region, education level, civil status, self-reported hypertension, self-reported high cholesterol and recency of blood pressure and cholesterol measurements. 
These variables are collected through the questionnaire and they are potential predictors for the lifestyle indicators and the participation.  
Covariates with missing data were imputed simultaneously with the main variables. The imputation models are specified as in Karvanen et al. \citep{karvanen2016selection}. In addition, we predict the number of hospitalizations for model checking purposes based on the same covariates as for daily smoking and heavy alcohol consumption.

\subsection{Evaluation of modelling assumptions}
We evaluate the modelling assumptions (1)-(3) using the background variables and the hospitalization history data. Assumption (1) is violated if there is an indication that either distribution of variables measured in the survey or distributions of background variables differ between participants and non-participants (including re-contact respondents). Assumption (2) does not hold if participants and non-participants (including re-contact respondents) differ with respect to their health indicators when conditioned on background variables. Assumptions (2) and (3) cannot be tested directly because there is no estimate of health indicator available for non-participants. Instead, they are evaluated by fitting a statistical regression model for the number of hospitalizations by each of the three groups and using the background variables as covariates.

We check if the hospitalization rates differ between the participants, the re-contact respondents and the non-participants. A difference is interpreted as an evidence of differences in the health indicator distributions. If the hospitalization rates for re-contact respondents can be assumed to be similar to non-participants' rates, then we can obtain information on the health of the non-participants from the re-contact data.

We utilize a zero-inflated negative binomial regression \citep{regressionmodelsr} as a model for the hospitalization data to evaluate assumptions (2) and (3). The zero-inflated model consists of two parts: the excess zeros model, and the model for the counts. The count model utilizes negative binomial distribution. The excess zero model describes the proportion of excess zeros (zero inflation) in addition to the zeros from the count model. Thus, a zero may occur from both of the models; the excess zero model or the count model.

We check the assumptions using full, five-year and one-year hospitalization history data. 
The longer the history, the more hospitalization events are expected. A high total number of events makes it easier to observe differences in the counts between the groups. However, as the health of an individual changes over time, hospitalization counts from a recent period may better describe the health at the time of the survey.


\section{Results}
The characteristics of collected survey data are described in Table \ref{tab-datadesc}. Among participants and re-contact respondents there are slightly more women than men. Among non-participants, the opposite is true, which indicates that women are more eager to participate. The mean age of non-participants is lower than the mean age of participants or re-contact respondents. 
The re-contact respondents seem to be less educated and more often single than the participants. For both men and women there are more smokers among re-contact respondents than among participants. For men, the proportion of heavy alcohol consumption is $6.8\%$ for both participants and re-contact respondents, but there is a lot of variation between the age groups. The proportion is much higher among the young re-contact respondents than among young participants. In age-group 25--34-years-old the proportion is exceptionally high ($15.9 \%$) compared to other groups. Among the re-contact respondents of the age groups 55--65 and 65--74 the proportion drops below the rates of participants. For women, in all age groups there is higher heavy alcohol consumption proportion for re-contacts than for participants.

Re-contact respondents seem to be more often smokers and heavy alcohol users than participants, except for heavy alcohol consumption among men where the prevalences are the same for participants and re-contact respondents. 

Table \ref{tab-nhospmodel} shows results for the assumption checking model. 
The risk of being hospitalized is higher for men than women and the risk increases with age.
A significant difference between participants and non-participants was observed for full, five-year and one-year hospitalization histories, while no difference between re-contact respondents and non-participants was found for five-year and one-year histories. These findings indicate that assumption (2) does not hold, while assumption (3) is supported.

Table \ref{tab-hospitalizations} presents the predicted hospitalization counts per 1000 individuals for each length of the hospitalization history. The proposed method, the MI-MNAR, has predicted counts which match the best with the observed full cohort counts. 	
This supports the assumption (3), which states that re-contacts represent the non-participants given their background variables. 
The match is more convincing for one-year and five-year histories than for full history. The hospitalization counts for participants, re-contact respondents and non-participants can be compared with each other. It is interesting to see that hospitalization counts per 1000 individuals are lower for female non-participants than for female re-contact respondents. For men, the contrary is true.

Table \ref{tab-comparison} describes the prevalences of daily smoking and heavy alcohol consumption estimated with different imputation models. MI-MNAR imputation results show that the point estimate of the prevalence of daily smoking for men is 28.5\%, which is 5.3 \%-points higher than what was measured from the participants only. For women, the corresponding imputed estimate is 19.0\%, which is 2.5 \%-points higher than the estimate based on the participants only. For smoking, the estimates from participants only do not lie within $95\%$ confidence interval of MI-MNAR imputations for men, and for women they are barely within the confidence interval. 
The point estimates by MI-MAR-NR are in all cases lower than the point estimates of MI-MAR and MI-MNAR.

The prevalence of heavy alcohol consumption for men by MI-MNAR method is 9.4\%. This is much higher than one would expect based on the heavy alcohol consumption rates of participants (6.8\%) and re-contact respondents (6.8\%). The key factors in the imputation of heavy alcohol consumption are smoking, sex, age, and region. Smoking strongly predicts heavy alcohol consumption in the estimated imputation model. Corresponding odds ratios for participants are 3.93 (2.87, 5.4) for men and 4.1 (2.57, 6.56) for women. Further, it can be seen from Table~1 that the heavy alcohol consumption is much more common among young re-contacts than among participants and non-participation is much more common among young people than among others. These facts together explain why MI-MNAR leads to high prevalence of heavy alcohol consumption in men.

\begin{table}
	\small\sf\centering
	\caption{The averages and proportions with their 95\% confidence intervals for background variables and health indicators.}
	\label{tab-datadesc}	
	\begin{center}
		\begin{tabular}{ llll }
			&  Participants  &  Re-contact  &  Non-participants \\ 
			&				 &  respondents	 & \\
			$N$ &  $5827$  &  $597$  &  $3576$  \\
			Women, \%  &  53.1 (51.5,54.6)  &  53.3 (48.6,58.0)  &  46.1 (44.0,48.1) \\ 
			Mean age, years  &  49.7 (49.3,50.0)  &  49.2 (48.1,50.3)  &  44.9 (44.4,45.3) \\ 
			~~~~Age group 25--34,  \%  &  18.7 (17.5,19.9)  &  21.3 (17.5,25.2)  &  30.5 (28.6,32.3) \\ 
			~~~~Age group 35--44, \%  &  18.0 (16.8,19.2)  &  15.8 (12.4,19.2)  &  21.7 (20.0,23.4) \\ 
			~~~~Age group 45--54, \%  &  22.1 (20.9,23.4)  &  22.8 (18.8,26.7)  &  20.1 (18.4,21.7) \\ 
			~~~~Age group 55--64, \%  &  23.7 (22.4,25.0)  &  26.2 (22.1,30.4)  &  17.7 (16.1,19.3) \\ 
			~~~~Age group 65--74, \%  &  17.5 (16.3,18.6)  &  13.9 (10.6,17.1)  &  10.1 (8.9,11.4) \\ 
			Education & & \\
			~~~~High, \%  &  37.6 (36.1,39.1)  &  34.7 (30.3,39.2)  &  -- \\ 
			~~~~Low, \%  &  30.9 (29.5,32.3)  &  34.8 (30.3,39.3)  &  -- \\ 
			Civil status &&\\
			~~~~Married, \%  &  52.4 (50.9,54.0)  &  49.8 (45.1,54.5)  &  -- \\ 
			~~~~Cohabiting, \%  &  18.6 (17.4,19.8)  &  17.1 (13.6,20.7)  &  -- \\ 
			~~~~Single, \%  &  15.4 (14.3,16.5)  &  19.3 (15.6,23.0)  &  -- \\ 
			~~~~Divorced, \%  &  10.7 (9.7,11.6)  &  11.4 (8.4,14.4)  &  -- \\ 
			~~~~Widow, \%  &  2.8 (2.3,3.4)  &  2.5 (1.0,3.9)  &  -- \\ 	 
			Daily smokers, men \%   &  23.2 (21.9,24.5)  &  28.5 (24.2,32.7)  &  -- \\ 
			~~~~Age group 25--34, \%  &  30.4 (29.0,31.8)  &  26.1 (21.9,30.2)  &  -- \\ 
			~~~~Age group 35--44, \%  &  24.4 (23.1,25.7)  &  36.3 (31.8,40.9)  &  -- \\ 
			~~~~Age group 45--54, \%  &  23.1 (21.9,24.4)  &  24.2 (20.2,28.2)  &  -- \\ 
			~~~~Age group 55--64, \%  &  24.3 (23.0,25.6)  &  31.4 (27.0,35.8)  &  -- \\ 
			~~~~Age group 65--74, \%  &  13.2 (12.2,14.3)  &  23.3 (19.3,27.2)  &  -- \\ 
			Daily smokers, women \% &  16.5 (15.3,17.6)  &  19.7 (15.9,23.4)  &  -- \\ 
			~~~~Age group 25--34, \%  &  21.0 (19.8,22.2)  &  16.3 (12.8,19.8)  &  -- \\ 
			~~~~Age group 35--44, \%  &  15.7 (14.6,16.9)  &  24.0 (19.9,28.0)  &  -- \\ 
			~~~~Age group 45--54, \%  &  17.5 (16.3,18.7)  &  19.6 (15.9,23.4)  &  -- \\ 
			~~~~Age group 55--64, \%  &  17.5 (16.3,18.7)  &  28.1 (23.9,32.3)  &  -- \\ 
			~~~~Age group 65--74, \%  &  9.2 (8.3,10.1)  &  4.7 (2.7,6.7)  &  -- \\ 
			Heavy alcohol users, men \% &  6.8 (6.1,7.6)  &  6.8 (4.4,9.2)  &  -- \\ 
			~~~~Age group 25--34, \%  &  5.9 (5.2,6.7)  &  15.9 (12.5,19.4)  &  -- \\ 
			~~~~Age group 35--44, \%  &  5.3 (4.6,6.0)  &  9.1 (6.4,11.8)  &  -- \\ 
			~~~~Age group 45--54, \%  &  9.6 (8.7,10.5)  &  5.4 (3.2,7.5)  &  -- \\ 
			~~~~Age group 55--64, \%  &  7.2 (6.4,8.0)  &  0.9 (0.0,1.8)  &  -- \\ 
			~~~~Age group 65--74, \%  &  5.3 (4.6,6.0)  &  1.6 (0.4,2.8)  &  -- \\ 
			Heavy alcohol users, women \% &  3.0 (2.5,3.5)  &  5.0 (3.0,7.1)  &  -- \\ 
			~~~~Age group 25--34, \%  &  4.3 (3.6,4.9)  &  6.8 (4.4,9.1)  &  -- \\ 
			~~~~Age group 35--44, \%  &  3.0 (2.5,3.5)  &  6.4 (4.1,8.7)  &  -- \\ 
			~~~~Age group 45--54, \%  &  3.8 (3.2,4.4)  &  5.1 (3.0,7.2)  &  -- \\ 
			~~~~Age group 55--64, \%  &  2.5 (2.0,3.0)  &  4.5 (2.6,6.5)  &  -- \\ 
			~~~~Age group 65--74, \%  &  1.0 (0.7,1.3)  &  1.5 (0.3,2.6)  &  -- \\
		\end{tabular}
	\end{center}
\end{table}

\begin{sidewaystable}
	\small\sf\centering
	\caption{Estimated parameters with their 95\% confidence intervals from the zero-inflated negative binomial regression model for the number of hospital visits. The model was fitted using three periods of history data; full history, five-year history and one-year. The reference levels for categorical variables sex and region are men and North Karelia.} 
	
	\label{tab-nhospmodel}	
	\begin{center}
		\begin{tabular}{lrlrlrl}
			&      \multicolumn{6}{c}{Estimate (95\% confidence interval)}\\\hline\\[-0.2cm]
			Count model     			& \multicolumn{2}{c}{Full history}			&		\multicolumn{2}{c}{Five years}  		& \multicolumn{2}{c}{One-year}\\\hline\\[-0.4cm]
			Intercept  				           & 0.84 & (0.71, 0.98)	&  -0.88 & (-1.19, -0.57)	&  -1.79 & (-2.49, -1.10)	\\
			Age: Men (10 years)  		    & 0.18 & (0.16, 0.20)	&   0.26 & (0.21, 0.31)		&   0.27 & (0.16, 0.39)	\\
			Age: Women (10 years)  	    & 0.33 & (0.30, 0.35)	&   0.22 & (0.17, 0.28)		&   0.08 & (-0.02, 0.18)	\\ 
			Sex (Woman)  				          &-0.51 & (-0.67, -0.34)	&   0.32 & (-0.05, 0.69)	&   1.22 & (0.35, 2.10)	\\ 
			Region: Northern Savonia	   & 0.00 & (-0.09, 0.09)	&  -0.03 & (-0.20, 0.13)	&   0.04 & (-0.23, 0.31)	\\ 
			Region: Turku and Loimaa    &-0.16 & (-0.25, -0.07)	&  -0.26 & (-0.43, -0.09)	&  -0.26 & (-0.55, 0.02)	\\ 
			Region: Helsinki and Vantaa &-0.30 & (-0.37, -0.22)	&  -0.45 & (-0.60, -0.31)	&  -0.46 & (-0.70, -0.23)	\\ 
			Region: Oulu  				         & 0.03 & (-0.05, 0.11)	&  -0.09 & (-0.24, 0.06)	&  -0.16 & (-0.41, 0.09)	\\ 
			Participant (Yes)  		      &-0.25 & (-0.30, -0.21)	&  -0.60 & (-0.71, -0.48)	&  -0.92 & (-1.14, -0.70)	\\ 
			Re-contact respondent (Yes)&-0.10 & (-0.19, -0.01)	&   0.02 & (-0.20, 0.24)	&   0.08 & (-0.33, 0.50)	\\\\ 
			Zero model &&&\\\hline\\[-0.4cm] 
			Intercept  				            &22.19 & (6.74, 37.63)	&   1.40 & (0.53, 2.26)		&   1.73 & (0.66, 2.80)	\\ 
			Age: Men (10 years)  		    &-9.23 & (-15.32, -3.15)	&  -0.56 & (-0.78, -0.34)	&  -0.31 & (-0.55, -0.07)	\\ 
			Age: Women (10 years)  	    &-1.46 & (-1.80, -1.11)	&  -0.44 & (-0.65, -0.22)	&  -0.42 & (-0.58, -0.26)	\\ 
			Sex (Women)  				          &-19.44 & (-34.93, -3.96)	&  -0.46 & (-1.78, 0.85)	&   0.99 & (-0.35, 2.33)	\\ 
			Participant (Yes)  		&-0.56 & (-1.11, 0.01)	&  -1.59 & (-2.66, -0.52)	&  -0.88 & (-1.37, -0.40)	\\ 
			Re-contact respondent (Yes) &-0.59 & (-1.96, 0.78)  	&   0.05 & (-0.58, 0.68)	&   0.12 & (-0.43, 0.67)
		\end{tabular}
	\end{center} 
\end{sidewaystable}

\begin{table}
	\small\sf\centering
	\caption{Hospitalisations per 1000 individuals by the length of the hospitalization history: using full history available, five-year history, one-year history. Four first rows describe the actual data and next three show the results of multiple imputations. The results of multiple imputations are to be compared with the numbers from full cohort.}
	\label{tab-hospitalizations}
	\begin{tabular}{ llll } 
		&\multicolumn{3}{c}{Estimate (95\% confidence interval)}\\\hline\\[-0.2cm]
		\textbf{Men:} 		&  Full history  	&  Five years  	 &  One-year\\\hline\\[-0.2cm] 
		Full cohort  		&  4305 (4126,4484) &  773 (718,829)  &  182 (163,201) \\ 
		~~~~Participants only  	&  3755 (3561,3948) &  647 (589,705)  &  147 (127,168) \\ 
		~~~~Re-contact resp.  &  4072 (3394,4751) &  941 (671,1212) &  227 (136,318) \\ 
		~~~~Non-participants  	&  5070 (4720,5420) &  919 (811,1027) &  223 (187,259) \\ 
		MI-MNAR  		&  4050 (3725,4374) &  834 (689,978)  &  188 (158,217) \\ 
		MI-MAR  		&  3784 (3526,4042) &  667 (602,731)  &  151 (130,171) \\ 
		MI-MAR-NR  &  3816 (3624,4008) &  675 (619,732)  &  152 (133,171) \\[0.2cm]
		\textbf{Women:}  	&   		    	&   	     	 &   \\\hline\\[-0.2cm] 
		Full cohort  		&  5445 (5237,5653) &  852 (783,921)  &  180 (160,200) \\ 
		~~~~Participants only  	&  5598 (5377,5818) &  799 (733,865)  &  156 (138,175) \\ 
		~~~~Re-contact resp.  		&  6514 (5581,7446) &  1073 (767,1379)&  269 (175,363) \\ 
		~~~~Non-participants  	&  5179 (4733,5625) &  918 (760,1076) &  207 (161,252) \\ 
		MI-MNAR  		&  5538 (5076,5999) &  929 (751,1108) &  222 (169,275) \\ 
		MI-MAR  		&  5168 (4967,5369) &  767 (692,842)  &  150 (132,168) \\ 
		MI-MAR-NR  &  5146 (4949,5343) &  760 (698,821)  &  153 (133,173) \\[0.2cm]
		\textbf{Both:}		&					&				 &   \\\hline\\[-0.2cm]
		Full cohort  		&  4880 (4742,5018) &  813 (769,857)  &  181 (167,195) \\ 
		~~~~Participants only  	&  4676 (4527,4825) &  723 (679,767)  &  152 (138,166) \\ 
		~~~~Re-contact resp.  		&  5293 (4696,5890) &  1007 (800,1214)&  248 (182,313) \\ 
		~~~~Non-participants  	&  5124 (4845,5403) &  919 (826,1012) &  215 (186,243) \\ 
		MI-MNAR  		&  4800 (4524,5076) &  882 (763,1001) &  205 (179,231) \\ 
		MI-MAR  		&  4482 (4320,4644) &  717 (667,767)  &  150 (137,163) \\ 
		MI-MAR-NR  &  4487 (4348,4626) &  718 (677,759)  &  152 (139,166) 
		
	\end{tabular}
\end{table}

\begin{table}
	\small\sf\centering
	\caption{Comparison of prevalence estimates of daily smoking and heavy alcohol consumption. The proposed method MI-MNAR is compared to alternative methods MI-MAR, MI-MAR-NR and estimates for the participants and re-contact respondents.}
	\label{tab-comparison}
	
	\begin{tabular}{ llll }
		&\multicolumn{2}{c}{Estimate (95\% confidence interval)}\\\hline\\[-0.2cm]		 		
		\textbf{Men:}  		&  Daily smokers (\%)	 & Heavy alcohol users (\%)\\\hline\\[-0.2cm]
		Participants  		&  23.2 (21.6,24.8)  & 6.8 (5.9,7.8)   \\
		Re-contact resp.  	&  28.5 (22.9,34.0)  & 6.8 (3.7,9.9)   \\
		MI-MNAR  		&  28.5 (25.9,31.2)  & 9.4 (7.2,11.6)  \\
		MI-MAR  		&  24.8 (23.1,26.5)  & 7.1 (5.7,8.4)   \\
		MI-MAR-NR  		&  23.7 (22.2,25.1)  & 6.7 (5.7,7.7)   \\[0.2cm]  
		\textbf{Women:}  	&  		  &   \\\hline\\[-0.2cm]  
		Participants  		&  16.5 (15.2,17.8)  & 3.0 (2.4,3.6)   \\
		Re-contact resp.  	&  19.7 (15.4,24.0)  & 5.0 (2.6,7.4)   \\
		MI-MNAR  		&  19.0 (15.8,22.2)  & 4.8 (3.4,6.3)   \\
		MI-MAR  		&  17.1 (15.6,18.5)  & 3.2 (2.4,3.9)   \\
		MI-MAR-NR  		&  16.5 (15.0,18.0)  & 3.1 (2.3,3.9)   \\[0.2cm]   
		\textbf{Both:}  	& 		  & \\\hline\\[-0.2cm]
		Participants  		&  19.6 (18.6,20.6)  & 4.8 (4.2,5.3)   \\
		Re-contact resp.  	&  23.7 (20.3,27.2)  & 5.9 (3.9,7.8)   \\
		MI-MNAR  		&  23.7 (21.5,25.9)  & 7.1 (5.6,8.6)   \\
		MI-MAR  		&  20.9 (19.7,22.0)  & 5.1 (4.4,5.8)   \\
		MI-MAR-NR  		&  20.1 (19.0,21.1)  & 4.9 (4.3,5.5)  
	\end{tabular}
\end{table}

\section{Discussion}

We studied the estimation of population-level health indicators from data that suffer from relatively high non-participation. The estimation utilized re-contact data, i.e. data from the non-participants who answered to a survey questionnaire when contacted again. With data from FINRISK 2012, we estimated the prevalence of daily smoking and heavy alcohol consumption using the MI-MNAR approach. These estimates were compared with the estimates obtained using less elaborated MI-MAR and MI-MAR-NR approaches and with the straightforward inclusion of participants only.

These approaches relied on different assumptions. The MI-MNAR approach assumed that re-contact respondents represent all non-participants when adjusted for the background variables, while the MI-MAR approaches used a stronger assumption that participants represent the whole population when adjusted for the background variables. The inclusion of the participants only (complete case analysis) used the strongest assumption that the participants represent the whole population.

The bias in the estimates depends on the validity of the assumptions. Many HESs report that participants and non-participants differ with respect to their health indicators, which violates the assumption of complete case analysis. This is also the case for the FINRISK 2012 data as the prevalence of daily smoking and heavy alcohol consumption for participants and re-contact respondents differ. We evaluated the other two assumptions using register-based history data about the hospitalizations of all people invited to the study. We checked if there were differences in numbers of hospitalizations between participants, re-contact respondents and the remaining non-participants when other variables were used as covariates.

We found out that if an individual had ever been hospitalized, the expected number of hospitalizations for re-contact respondents and non-participants were the same. 
In addition, we predicted the number of hospitalizations using three multiple imputation approaches. We observed that the predictions from the MI-MNAR approach matched best with the true number of hospitalizations. These findings support the assumption which is utilized by MI-MNAR approach.

The evaluation of assumptions (2) and (3) was based on the idea that the number of hospitalizations is associated with the health status. If the number of hospitalizations differs between re-contact respondents and non-participants, then there is likely to be a difference in distributions of health indicators between the groups. Otherwise, the distributions are assumed to be the same. As we used the hospitalizations before the study, the symptoms are not caused by the health condition during the survey date but are associated with them. 

This makes us to think that the hospitalizations before the study are a less convincing source of evidence than prospective follow-up data that have been earlier used to evaluate the assumptions for FINRISK 2007 \citep{karvanen2016selection}. 
If the follow-up data are available, then we recommend using them \cite{karvanen2016selection}. Otherwise, we recommend using the proposed method instead of not checking the assumptions at all.
Differently from prospective follow-up data, the hospitalization history data are readily available shortly after the study.
In principle, hospitalization history data could be used directly in the imputation model such that instead of just evaluating the assumptions (1)--(3) the imputations would be predicted based on the hospitalization history data. How to optimally do this and what is the benefit are questions to be further investigated.

The setup for FINRISK 2012 was similar to FINRISK 2007, which allows a comparison between the studies. Using the data from the participants only, the point estimates for smoking prevalence were 21.8\% in 2007 \citep{karvanen2016selection} and 19.6\% in 2012. Similarly, the prevalences of heavy alcohol consumption were estimated as 5.2\% in 2007 and 4.8\% in 2012. Thus, based on the participants only, there seems to be a positive development. 

The situation looks different if MI-MNAR approach is used. Then the estimated prevalence of daily smoking appears as $27.1$\% in 2007 \citep{karvanen2016selection} and $23.7$\% in 2012. Estimated prevalences of heavy alcohol consumption are $6.8$\% in 2007 and $7.1$\% in 2012. 
Thus, there seem to be notable differences in the prevalence estimates between the approaches. The MI-MNAR approach produces the widest confidence intervals in comparison to MI-MAR, MI-MAR-NR and participants approach, all of which are based on unrealistic assumptions.

As noted by many authors \citep{karvanen2016selection,van2003survey,jousilahti2005total,harkanen2016systematic,tolonen2005effect,nummela2011aging}, missing data caused by non-participation is a serious problem in HES. Our results support the idea that re-contact data can improve the reliability of the health indicators of non-participants and provide information about the selectivity.

Although the assumption for MI-MNAR holds for FINRISK 2012 data, it may not hold for other HESs. For example, LEIDEN 85-plus study \citep{bootsma2002high} observed that mortality risk of re-contact respondents was similar to participants for old persons. In such a situation, re-contact data were not useful for bias reduction. As the populations of FINRISK and LEIDEN 85-plus differ a lot, the results are not directly comparable.

According to our knowledge, re-contact data have only occasionally been collected in HESs.  Our results suggest that re-contact data can provide information about the health indicators of non-participants and selectivity of non-participation. 
Therefore, we recommend that HESs would collect re-contact questionnaire data and that the same self-reported questions would be asked for participants to allow comparison.

Obtaining representative estimates about sensitive health indicators associated with selective non-participation is important for data-driven decision making in national health policy. Our work shows that re-contact data have potential to help to reduce the selection bias.
When used together with hospitalization register data, the assumptions for which the estimation of population-level health indicators is based on can be evaluated soon after the survey. 

\section*{Declaration of conflicting interests}
The Authors declare that there is no conflict of interest.

\section*{Funding}
This work was supported by Academy of Finland [grant number 266251].

\bibliographystyle{apalike}

\end{document}